\documentclass[11pt,a4paper]{article}
\usepackage{amsmath}
\usepackage{bm}
\usepackage{color}
\RequirePackage[numbers,sort&compress]{natbib}
\usepackage[colorlinks=true,urlcolor=blue,linkcolor=blue,citecolor=red]{hyperref}
\usepackage[]{hyperref}
\renewcommand{\thefootnote}{\fnsymbol{footnote}}

\def\onehalf{{\textstyle{\frac{1}{2}}}}

\def\be{\begin{equation}}
\def\ee{\end{equation}}
\def\ba{\begin{eqnarray}}
\def\ea{\end{eqnarray}}
\begin{document}

\begin{center}
{\Large \bf De Sitter-Invariant Black Holes}
\vskip 0.8cm
{\bf Diego F. L\'opez,$^{a}$
Salman Abarghouei Nejad,$^{b}$\footnote{Current~address: Department of Physical Sciences, Embry-Riddle Aeronautical University, Day\-tona Beach, FL 32114, USA.}
Jos\'e G. Pereira$^{b,}$\footnote{Correspondence: jg.pereira@unesp.br}}

\vskip 0.2cm
$^a${\it Department of Mathematics and Statistics\\ 
Dalhousie University\\
Halifax, Canada} 
\vskip 0.2cm
$^b${\it Universidade Estadual Paulista \\ 
Instituto de F\'{\i}sica Te\'orica \\
S\~ao Paulo, Brazil}

\vskip 0.8cm

\begin{quote}
{\bf Abstract:}~{\small Unlike the standard Poincaré-invariant special relativity, the de Sitter-invariant special relativity incorporates proper conformal transformations in the spacetime kinematics. Consequently, in addition to the usual event horizon, sourced by the energy-momentum current, the de Sitter-invariant black hole also features a de Sitter cosmic horizon, sourced by the proper conformal current. Therefore, besides lodging matter energy, it also lodges dark energy. Due to this additional structure, which originates from the proper conformal sector of the theory, such a black hole can establish a connection between the dynamics of black holes and the universe's evolution.}
\end{quote}

\end{center}
\vskip 1.0cm


\renewcommand{\thefootnote}{\arabic{footnote}}
\setcounter{footnote}{0}
\section{Introduction}
\label{intro}

A fundamental issue of the standard Poincaré-invariant approach is the absence of a special relativity governing the Planck scale kinematics. Even though Einstein's special relativity successfully describes the kinematics of particles moving at velocities approaching the speed of light, it fails to describe the kinematics of particles at energies approaching the Planck energy~\cite{AC1, AC2}.

To clarify this point, let us recall that local (or proper) conformal symmetry is a broken symmetry of nature, which is expected to become exact at the Planck scale~\cite{coleman}. The problem is that Einstein's special relativity does not include proper conformal transformations in the spacetime kinematics, making it unclear how this symmetry could become relevant at the Planck scale~\cite{tHooft}. To address this issue, one must seek a new special relativity that incorporates proper conformal transformations into the kinematics of spacetime. 

A natural solution to this problem is to replace the Poincaré-invariant special relativity with the de Sitter-invariant special relativity~\cite{dSsr0}. Since the de Sitter group arises from the Poincaré group by substituting translations with a combination of translations and proper conformal transformations~\cite{dSgeod}, upon  the above replacement one is effectively incorporating proper conformal transformations into spacetime kinematics. Notably, this incorporation does not change the dimension of the kinematic group, as both Poincaré and de Sitter are ten-dimensional groups. 

On account of the inclusion of proper conformal transformations into spacetime kinematics, in contrast to Einstein's special relativity the de Sitter-invariant special relativity provides a consistent description of the Planck scale kinematics. Accordingly, within the domain of the de Sitter invariant special relativity, all relativistic theories will be locally invariant under the de Sitter group. For this reason, this formulation of relativistic physics will be referred to as the de Sitter-invariant approach.

When the Poincaré-invariant special relativity is substituted with the de Sitter-invariant special relativity, all relativistic theories undergo concomitant changes. In particular, the Poincaré-invariant general relativity transforms into the de Sitter-invariant general relativity~\cite{DE}, with the corresponding field equation referred to as the de Sitter-invariant Einstein's equation.

In the Poincaré-invariant general relativity, all solutions to the gravitational field equation are spacetimes that reduce locally to Minkowski, where the local kinematics governed by the Poincaré-invariant special relativity group takes place. Conversely, in the de Sitter-invariant general relativity, all solutions to the gravitational field equations are spacetimes that reduce locally to de Sitter, where the local kinematics governed by the de Sitter-invariant special relativity takes place.\footnote{Spacetimes that do not reduce locally to Minkowski have been known for a long time and come under the name of Cartan geometry~\cite{CartanGeometry}.} 

By construction, the de Sitter-invariant approach becomes relevant for energies comparable to the Planck energy and at the universe's large scale. For this reason, it is essential for the study of quantum gravity and cosmology. Since black holes also involve extremely high energies, they must be obtained as solutions to the de Sitter-invariant Einstein's equation. The black hole solution obtained from the Poincaré-invariant Einstein's equation is incomplete because it lacks the conformal sector introduced by the de Sitter group.

Using the de Sitter-invariant approach, the primary purpose of this paper is to obtain the de Sitter-invariant black hole solution and discuss its main properties and differences concerning the Poincaré-invariant black hole.

\section{Minkowski and de Sitter as quotient spaces}
\label{MinkodS}

Spacetimes with constant sectional curvature are maximally symmetric because they carry the maximum number of Killing vectors. Flat Minkowski spacetime {$M$} is the simplest example. Its kinematic group is the Poincaré group ${\mathcal P} = {\mathcal L} \oslash {\mathcal T}$, the semi-direct product of the Lorentz group ${\mathcal L}$ and the translation group ${\mathcal T}$. Algebraically, it is defined as the quotient space~\cite{GeoLivro}
\begin{equation}\nonumber
M = {\mathcal P}/{\mathcal L}.
\end{equation}

The Lorentz subgroup is responsible for the isotropy around a given point of $M$, and the translation symmetry enforces this isotropy around all other points. In this case, homogeneity means that all points of Minkowski are equivalent under spacetime translations. One then says that Minkowski is {\em transitive} under translations, whose generators are written~as
\begin{equation}
P_\nu = \delta^\alpha_\nu \partial_\alpha,
\label{transiM}
\end{equation}
with $\delta^\alpha_\nu$ the Killing vectors of translations.

The de Sitter spacetime is also maximally symmetric. Its kinematic group is the de Sitter group~$SO(1,4)$. Consequently, it can be defined as the quotient space~\cite{GeoLivro}
\begin{equation}
dS = SO(1,4) / {\mathcal L}.
\end{equation}

Similarly to Minkowski, the Lorentz subgroup is responsible for the isotropy around a given point of $dS$. However, the de Sitter homogeneity differs from Minkowski's. To determine it, we recall that the de Sitter spacetime can be viewed as a hyperboloid embedded in the $(1+4)$-dimensional pseudo-Euclidean space with Cartesian coordinates $\chi^A$ ($A, B, C \dots = 0, \dots 4$) and metric
\begin{equation}\nonumber
\eta_{AB} = {\rm diag}\, (+1,-1,-1,-1,-1),
\end{equation}
 whose points satisfy~\cite{HE}:
\begin{equation}
\eta_{AB} \, \chi^A \chi^B = - \, l^2.
\label{constraint0}
\end{equation}

In terms of the embedding coordinates~${\chi^A}$, the generators of infinitesimal de Sitter transformations are written in the form
\begin{equation}
L_{A B} \equiv \xi^{\;C}_{AB} \, \frac{\partial}{\partial \chi^C},
\label{dsgene}
\end{equation}
with
\begin{equation}
\xi^{\;C}_{AB} = \eta_{AD} \chi^D \, \delta^C_B - 
\eta_{BD} \chi^D \, \delta^C_A  
\end{equation}
the corresponding Killing vectors. 

In terms of the four-dimensional stereographic coordinates $\{x^\mu\}$~\cite{StereoCoordi}, the ten de Sitter generators \eqref{dsgene} split into
\begin{equation}
L_{\mu \nu} = \xi^{\,\alpha}_{\mu \nu} \, \partial_\alpha \qquad \mbox{and} \qquad
{\Pi}_\nu \equiv \frac{L_{4 \nu}}{l} = \xi^\alpha_\nu \,\partial_\alpha,
\label{dSgene4dim} 
\end{equation}
where $L_{\mu \nu}$ represent the Lorentz generators and ${\Pi}_\nu$ are the generators defining the de Sitter homogeneity, with $\xi^{\,\alpha}_{\mu \nu}$ and $\xi^\alpha_\nu$ the corresponding Killing vectors. 
In these coordinates, the Killing vectors associated with the de Sitter homogeneity are written in the form
\begin{equation}
\xi^\alpha_\nu =
\delta^\alpha_\nu - \frac{1}{4 l^2} \, \vartheta^\alpha_\nu,
\label{dsKilling3Bis}
\end{equation}
where
\begin{equation}
\delta^\alpha_\nu \quad \mbox{and} \quad \vartheta^\alpha_\nu = 2\, \eta_{\nu \rho} \, x^\rho x^\alpha - 
\sigma^2 \delta^\alpha_\nu
\end{equation}
are the Killing vectors of translations and proper conformal transformations, with 
\[
\sigma^2 = \eta_{\alpha \beta} \, x^\alpha x^\beta.
\]
In this case, the generators~${\Pi}_\nu$ can be rewritten as
\begin{equation}
{\Pi}_\nu = {P}_\nu - \frac{1}{4 l^2}\, {K}_\nu,
\label{pi3}
\end{equation}
where
\begin{equation}
{P}_\nu = \delta^\alpha_\nu \, \partial_\alpha \quad \mbox{and} \quad
{K}_\nu = \vartheta^\alpha_\nu \, \partial_\alpha
\label{TransGenerators}
\end{equation}
are the translation and proper conformal generators~\cite{ConfCurrent}. According to the generators~\eqref{pi3}, all points of de Sitter are equivalent under a combination of translations and proper conformal transformations~\cite{ccc}, known as de Sitter translations.\footnote{The generators $\Pi_\nu$ are not really translations but pseudo-rotations in the planes $(4,\nu)$.}

Equations~\eqref{transiM} and~\eqref{pi3} show that, whereas Minkowski is transitive under translations, the de Sitter spacetime is transitive under a combination of translations and proper conformal~transformations. As we will see in the next section, this difference plays a crucial role in deducing the gravitational field equations.

\section{Gravitational field equations}
\label{dSmGR}

We now outline how the Poincaré and de Sitter-invariant Einstein's equations are obtained. 

\subsection{Poincar\'e-invariant Einstein's equation}

In standard Poincaré-invariant general relativity, all solutions to the gravitational field equation are spacetimes that reduce locally to Minkowski. Since Minkowski is transitive under translations, a diffeomorphism in these spacetimes is defined as a local translation,
\begin{equation}
\delta_P x^\mu = \delta^\mu_\alpha \, \epsilon^\alpha(x),
\label{OrDiff}
\end{equation}
with $\delta^\mu_\alpha$ the Killing vectors of translations. From Noether's theorem, the invariance of the source Lagrangian under the diffeomorphism~\eqref{OrDiff} yields the conservation law
\begin{equation}
\nabla_\nu T^{\mu \nu} = 0,  \qquad T^{\mu \nu} = \delta^{\mu}_{\alpha} \, T^{\alpha \nu},
\end{equation}
with the conserved current given by the projection of the energy-momentum current along the Killing vectors of translations.
Accordingly, Einstein's equation is written as
\begin{equation}
G^{\mu \nu} \equiv R^{\mu \nu} - {\textstyle{\frac{1}{2}}} g^{\mu \nu} R = - \frac{8 \pi G}{c^4} \,T^{\mu \nu}.
\label{OrdiEinstein}
\end{equation}

\subsection{De Sitter-invariant Einstein's equation}

In de Sitter-invariant general relativity, all solutions to the gravitational field equations are spacetimes that reduce locally to de Sitter, where the local kinematics takes place. Since de Sitter is transitive under de Sitter translations, a diffeomorphism in these spacetimes is defined as a local de Sitter translation~\cite{dSgeod}
\begin{equation}
\delta_{\Pi} x^\mu = \xi^{\mu}_{\alpha} \, \epsilon^{\alpha}(x),
\label{dStrans}
\end{equation}
with $\xi^{\mu}_{\alpha}$ the corresponding Killing vectors. From Noether's theorem, the invariance of the source Lagrangian under the diffeomorphism~\eqref{dStrans} yields the covariant conservation law~\cite{DE}
\begin{equation}
\nabla_\nu \Pi^{\mu \nu} = 0, \qquad \Pi^{\mu \nu} = \xi^{\mu}_{\alpha} \, T^{\alpha \nu},
\end{equation}
with the conserved current given by the projection of the energy-momentum current along the Killing vectors of de Sitter translations.
Accordingly, the de Sitter-invariant Einstein's equation is written as
\begin{equation}
{\mathcal G}^{\mu \nu} \equiv {\mathcal R}^{\mu \nu} - {\textstyle{\frac{1}{2}}} g^{\mu \nu} {\mathcal R} = - \frac{8 \pi G}{c^4} \, \Pi^{\mu \nu},
\label{NewModiEinstein}
\end{equation}
where the Riemann tensor ${\mathcal R}^\alpha{}_{\beta \mu \nu}$ represents both the dynamic curvature of general relativity and the kinematic curvature of the background de Sitter spacetime. In this theory, therefore, the cosmological term~$\Lambda$ is naturally encoded in the background de Sitter spacetime, and does not need to be added by hand to the gravitational field equation. Consequently, $\Lambda$ does not appear explicitly in the field equation, and the second Bianchi identity does not restrict it to be constant. This is a distinctive property of the de Sitter-invariant approach to gravitation and cosmology. 
 
\subsection{The source of the cosmological term}
\label{dSsource}

Substituting the Killing vectors~\eqref{dsKilling3Bis} in the source current~$\Pi^{\mu \nu}$, it splits in the form
\begin{equation}
\Pi^{\mu \nu} = T^{\mu \nu} - ({1}/{4 l^2})\, K^{\mu \nu},
\label{Pi=T-K}
\end{equation}
where
\begin{equation}
T^{\mu \nu} = \delta^\mu_\alpha \, T^{\alpha \nu}\qquad \mbox{and} \qquad K^{\mu \nu} = \vartheta^\mu_\alpha \, T^{\alpha \nu}
\label{T&K}
\end{equation}
are the energy-momentum and the proper conformal currents. 
Analogously to the source decomposition, the Einstein tensor ${\mathcal G}^{\mu \nu}$ splits in the form
\begin{equation}
{\mathcal G}^{\mu \nu} = G^{\mu \nu} - \hat{G}^{\mu \nu}, 
\end{equation}
where ${G}^{\mu \nu}$ is the Einstein tensor of general relativity tensor and $\hat{G}^{\mu \nu}$ is the Einstein tensor of the background de Sitter spacetime. In this case, the de Sitter-invariant Einstein's equation~\eqref{NewModiEinstein} takes the form\footnote{The Einstein tensor $\hat G^{\mu \nu}$ is actually written in terms of the spacetime metric. Only locally, where spacetime reduces to de Sitter, it becomes the Einstein tensor of the background de Sitter spacetime.}
\begin{equation}
G^{\mu \nu} - \hat G^{\mu \nu}
= - \frac{8 \pi G}{c^4} \Big[T^{\mu \nu} - (1/4l^2)\, K^{\mu \nu} \Big].
\label{NewEinstein2}
\end{equation}

We see from this equation that any matter source in locally de Sitter spacetimes gives rise to both an energy-momentum and a proper conformal current. The energy-momentum current $T^{\mu \nu}$ retains its role as the source of gravitation. In contrast, the proper conformal current $K^{\mu \nu}$ appears as the source of the local de Sitter spacetime, whose sectional curvature represents the cosmological term~$\Lambda$. 

Due to the different nature of the source currents, whereas the gravitational part of the equation is essentially dynamic, the conformal part is purely algebraic. This property is consistent with the non-propagating character of the cosmological term~$\Lambda$. Note furthermore that ordinary matter is the source of both gravitation and dark energy. In the next section, we discuss how this peculiar property emerges from the de Sitter-invariant approach.

\subsection{Repulsive interaction from ordinary matter}
\label{NegPressure}
Taking the trace and identifying $\hat{G}^\mu{}_\mu = - \Lambda$, Eq.~\eqref{NewEinstein2} assumes the form
\begin{equation}
R - \Lambda = \frac{8 \pi G}{c^4} \Big[ T^\mu{}_\mu - ({1}/{4 l^2})\, K^\mu{}_\mu \Big].
\label{NewEinsteinTrace} 
\end{equation}
We see from this equation that, whereas the source of the scalar curvature~$R$ is the trace of the energy-momentum current, the source of~$\Lambda$ is the trace of the proper conformal current.

We now consider a perfect fluid whose energy-momentum tensor in co-moving coordinates is written as
\begin{equation}
T^\mu{}_\nu = {\rm diag} \left(\varepsilon_m, - p_m, - p_m, - p_m \right),
\label{CC1}
\end{equation}
where $\varepsilon_m$ and $p_m$ are the matter energy density and pressure, respectively. Its trace takes the form
\begin{equation}
T^{\mu}{}_{\mu} \equiv \delta^\mu_\alpha T^\alpha{}_\mu = \varepsilon_m - 3 p_m.
\label{TraceEM}
\end{equation}
On the other hand, the trace of the proper conformal current is written as
\begin{equation}
K^{\mu}{}_\mu = \vartheta^\mu_\alpha T^\alpha{}_\mu \equiv
\big(2\, \eta_{\alpha \rho} x^\rho x^\mu - \sigma^2 \delta^\mu_\alpha \big) T^\alpha{}_\mu.
\label{Ktrace}
\end{equation}
As an explicit computation shows, we obtain~\cite{dScosmo}
\begin{equation}
\frac{K^\mu{}_\mu}{4 l^2} = \varepsilon_\Lambda + 3 p_\Lambda.
\label{KtraceBis2}
\end{equation}
Using Eqs.~\eqref{TraceEM} and \eqref{KtraceBis2}, the field equation~\eqref{NewEinsteinTrace} can be rewritten in the form
\begin{equation}
{R} - \Lambda = \frac{8 \pi G}{c^4} \Big[\big(\varepsilon_m - 3 p_m \big) -
\big(\varepsilon_\Lambda + 3 p_\Lambda \big) \Big].
\label{NewEinsteinTrace2}
\end{equation}

Since ordinary matter is the source of both gravitation and dark energy, the corresponding source fields satisfy the same equations of state,
\begin{equation}
p_m = w\, \varepsilon_m \qquad \mbox{and} \qquad p_\Lambda = w\, \varepsilon_\Lambda,
\label{EquState}
\end{equation}
with $w$ a numerical constant. 
In spite of this fact, the pressures~$p_m$ and $p_\Lambda$ naturally enter the field equation~\eqref{NewEinsteinTrace2} with opposite signs relative to $\varepsilon_m$ and $\varepsilon_\Lambda$, respectively. This difference arises solely from the mathematical intricacies of the proper conformal current, the source of the background de Sitter spacetime. Consequently, dark energy naturally emerges as a repulsive force.\footnote{For a detailed discussion of this point, see Ref.~\cite{dScosmo}, Section~3.5.}
\section{Schwarzschild solutions}

Following the standard procedure, we now obtain the de Sitter-invariant Schwarzschild solution. For comparison, we first obtain the Poincaré-invariant Schwarzschild solution.

\subsection{Spherically symmetric ansatz}

Let us consider a centrally symmetric distribution of matter. Since the gravitational field produced by such a source will also have central symmetry, its metric tensor can be written in the form~\cite{LandauLifshitz}:
\begin{equation}
ds^2 = e^\kappa \, c^2 \, dt^2 - e^\lambda \, dr^2 - r^2 \left(d\theta^2 + 
\sin^2\theta d\phi^2 \right),
\label{Schwar1}
\end{equation}
with $\kappa(r, t)$ and $\lambda(r, t)$ functions of the coordinates $r$ and $t$. Denoting by $\{x^0, x^1, x^2, x^3 \}$, respectively, the coordinates $\{ct, r, \theta, \phi \}$, the non-zero components of the metric tensor are:
\begin{equation}
g_{00} = e^\kappa , \qquad g_{11} = - e^\lambda , \qquad g_{22} = - r^2 , \qquad g_{33} = - r^2 \, \sin^2 \theta.
\end{equation}

The inverse components are:
\begin{equation}
g^{00} = e^{-\kappa} , \qquad g^{11} = - e^{-\lambda} , \qquad g^{22} = - r^{-2} , \qquad g^{33} = - r^{-2} \, \sin^{-2} \theta.
\end{equation}

Denoting differentiation with respect to $r$ with a `prime' and differentiation with respect to $ct$ with a `dot,' the non-vanishing components of the Levi-Civita connection are:
\begin{equation}
\begin{array}{lll}
\Gamma^1{}_{11} = {\lambda'}/{2} & \Gamma^0{}_{10} = {\kappa'}/{2} & \Gamma^2{}_{33} = - \sin \theta \, \cos \theta \\
{} & {} & \\
\Gamma^0{}_{11} = ({\dot \lambda}/{2})\, e^{\lambda - \kappa} & \Gamma^1{}_{22} = - r e^{-\lambda} &\Gamma^1{}_{00} = ({\kappa'}/{2})\, e^{\kappa - \lambda} \\
{} & {} & \\
\Gamma^2{}_{12} = \Gamma^3{}_{13} = {1}/{r} & \Gamma^3{}_{23} = \cot \theta & \Gamma^0{}_{00} = {\dot \kappa}/{2} \\
{} & {} & \\
\Gamma^1{}_{10} = {\dot \lambda}/{2} & \Gamma^1{}_{33} = - r \sin^2 \theta \, e^{- \lambda}. &
\end{array} 
\label{LC}
\end{equation}

\subsection{Poincar\'e-invariant Schwarzschild solution}

For comparison with the de Sitter-invariant case, we first review how the standard Schwarzschild solution is obtained from the Poincaré-invariant Einstein's field equation
\begin{equation}
{R}^\mu{}_\nu - 
\onehalf \delta^\mu_\nu\, R = - \frac{8 \pi G}{c^4} T^\mu{}_\nu.
\label{EinEq12}
\end{equation}
Computing the Ricci tensor ${R}^\mu{}_\nu$ of the Levi-Civita connections~\eqref{LC} and substituting in Einstein's equation~\eqref{EinEq12}, a straightforward computation yields the equations:
\begin{equation}
- e^{-\lambda} \left(\frac{\kappa'}{r} + \frac{1}{r^2} \right) + \frac{1}{r^2} = -
\frac{8 \pi G}{c^4} \, T^1{}_1
\label{112}
\end{equation}
 
\begin{eqnarray}
- {\onehalf} e^{-\lambda} \bigg(\kappa'' + \frac{\kappa'^2}{2} + \frac{\kappa' - \lambda'}{r} - \frac{\kappa' \lambda'}{2} \bigg) + 
{\onehalf} e^{-\kappa} \bigg(\ddot \lambda + \frac{\dot \lambda^2}{2} - \frac{\dot \lambda \dot \kappa}{2} \bigg) = - \frac{8 \pi G}{c^4} \, T^2{}_2 =  
- \frac{8 \pi G}{c^4} \, T^3{}_3
\label{212}
\end{eqnarray}

\begin{equation}
- e^{-\lambda} \bigg(\frac{1}{r^2} - \frac{\lambda'}{r} \bigg) + \frac{1}{r^2} = -
\frac{8 \pi G}{c^4} \, T^0{}_0
\label{312}
\end{equation}
\begin{equation}
- e^{-\lambda} \, \frac{\dot \lambda}{r} = - \frac{8 \pi G}{c^4} \, T^1{}_0.
\label{412}
\end{equation}
All other components of the field equation~\eqref{EinEq12} vanish identically. 

We look for a vacuum solution, i.e., a solution that holds outside the masses producing the gravitational field, where the source current $T^\alpha{}_\nu$ vanishes. 
After setting the source current $T^\alpha{}_\nu$ equal to zero, and considering that Eq.~\eqref{212} is redundant because it follows from the other three equations, we are left with
\begin{equation}
e^{-\lambda} \left(\frac{\kappa'}{r} + \frac{1}{r^2} \right) - \frac{1}{r^2} = 0
\label{512}
\end{equation}
\begin{equation}
~e^{-\lambda} \bigg(\frac{\lambda'}{r} - \frac{1}{r^2} \bigg) + \frac{1}{r^2} = 0
\label{612}
\end{equation}
\begin{equation}
\dot \lambda = 0.
\label{712}
\end{equation}

Equation~\eqref{712} shows that $\lambda$ does not depend on the time. Furthermore, adding \linebreak \mbox{Eqs.~\eqref{512} and~\eqref{612}} we find $\lambda' + \kappa' = 0$, whose integration yields
\begin{equation}
\lambda + \kappa = a(t)
\label{ft12}
\end{equation}
with $a(t)$ an arbitrary function of time. However, when we chose the quadratic interval $ds^2$ in the form \eqref{Schwar1}, there remained the possibility of an arbitrary transformation of time of the form $t = a(t')$. Such a transformation is equivalent to adding to $\kappa$ an arbitrary function of the time. This process allows us to make $a(t)$ in \eqref{ft12} vanish. Without loss of generality, we can thus set
\begin{equation}
\lambda + \kappa = 0.
\label{ft12bis}
\end{equation}

Equation~\eqref{612} can be integrated and yields
\begin{equation}
g_{00} \equiv e^{-\lambda} \equiv e^\kappa = 1 + \frac{\beta}{r},
\label{g00Mink}
\end{equation}
where $\beta$ is an integration constant. To determine this constant, we recall that far from the gravitating body, the gravitational field is weak and can be represented by its Newtonian limit, in which case we have the relation
\begin{equation}
g_{00} = 1 + 2 \, \frac{\phi}{c^2}
\label{ge0012}
\end{equation}
with $\phi$ the Newtonian potential
\begin{equation}
\phi = - \frac{GM}{r}.
\label{Newton12}
\end{equation}

Comparing equations~\eqref{g00Mink} and~\eqref{ge0012}, the constant $\beta$ is found to be
\begin{equation}
\beta = - r_M \equiv - \frac{2 G M}{c^2},
\label{SchwarRadius}
\end{equation}
with $r_M $ the radius of the black hole event horizon. The metric~\eqref{Schwar1} can then be recast in the form
\begin{equation}
ds^2 = f(r) \, c^2 \, dt^2 - 
\frac{dr^2}{f(r)} - r^2 \left(d\theta^2 + \sin^2\theta d\phi^2 \right)
\label{Schwar2}
\end{equation}
with:
\begin{equation}
f(r) = 1 - \frac{r_M}{r}.
\end{equation}

This is the usual Schwarzschild solution. Since
\begin{equation}\nonumber
\lim_{r \to \infty} {f(r)} = 1,
\end{equation}
the Schwarzschild metric reduces to Minkowski at infinity.

\subsection{De Sitter-invariant Schwarzschild solution}
\label{DSinvBH}

We now obtain the Schwarzschild solution to the de Sitter-invariant Einstein's field equation 
\begin{equation}
{\mathcal R}^\mu{}_\nu - {\textstyle{\frac{1}{2}}} \delta^\mu{}_\nu {\mathcal R} = - \frac{8 \pi G}{c^4} \, \Pi^\mu{}_\nu.
\label{NewModiEinsteinBIS}
\end{equation}
In contrast to Poincaré-invariant Einstein's field equation, whose source is the energy-momentum current~$T^\mu{}_\nu$, the source of the de Sitter-invariant Einstein's field equation is a combination of energy-momentum and proper conformal currents:
\be
\Pi^\mu{}_\nu = T^\mu{}_\nu -  \frac{K^\mu{}_\nu}{4l^2}.
\ee
This means that the de Sitter-invariant field equations can be obtained from the Poincaré-invariant field equations~(\ref{112}-\ref{412}) by replacing
\be
T^\mu{}_\nu \to \Pi^\mu{}_\nu = T^\mu{}_\nu - \frac{K^\mu{}_\nu}{4 l^2}.
\label{Replace}
\ee
However, since black holes involve high energy densities, the local value of $\Lambda$ will be huge, corresponding to a tiny de Sitter pseudo-radius~$l$. In this case, the proper conformal current becomes dominant concerning the energy-momentum current, and the replacement~\eqref{Replace} takes the form
\be
T^\mu{}_\nu \to -  \frac{K^\mu{}_\nu}{4l^2}.
\label{Replace2}
\ee
Furthermore, besides depending on the coordinate $r$, the parameters $\kappa$ and $\lambda$ also depend on the pseudo-radius~$l$ of the local de Sitter spacetime. Accordingly, the de Sitter-invariant version of equations~\eqref{512}-\eqref{712} are written as
\begin{equation}
e^{-\lambda(l)} \left(\frac{\kappa'(l)}{r} + \frac{1}{r^2} \right) - \frac{1}{r^2} = - \frac{8 \pi G}{c^4} \, \frac{K^1{}_1}{4l^2}, 
\label{512bis}
\end{equation}
\begin{equation}
~e^{-\lambda(l)} \left(\frac{\lambda'(l)}{r} - \frac{1}{r^2} \right) + \frac{1}{r^2} = + \frac{8 \pi G}{c^4} \, \frac{K^0{}_0}{4l^2},
\label{612bis}
\end{equation}
\begin{equation}
\dot \lambda(l) = \frac{8 \pi G}{c^4} \, \frac{K^1{}_0}{4l^2}.
\label{712bis}
\end{equation}

One should remark that since $K^\mu{}_\nu$ is the source of the background de Sitter spacetime, not of gravitation, its presence does not spoil the vacuum condition. The presence of the proper conformal current in the field equations~(\ref{512bis}-\ref{712bis}) is crucial for the local de Sitter invariance of the black hole. When the proper conformal current can be neglected, the background de Sitter spacetime reduces to Minkowski, and we obtain the usual Poincaré-invariant black hole solution, which does not carry dark energy.

For symmetry reasons, the component $K^1{}_0$ of the proper conformal current vanishes, yielding the constraint $\lambda = -\kappa$. Consequently, one obtains
\begin{equation}
\frac{8 \pi G}{c^4} \,\frac{K^0{}_0}{4l^2} = \frac{8 \pi G}{c^4} \,\frac{K^1{}_1}{4l^2} \equiv F(r,l),
\end{equation}
with $F(r,l)$ an arbitrary function to be determined by the boundary conditions. \linebreak \mbox{Eqs. \eqref{512bis} and \eqref{612bis}} can then be rewritten as:
\begin{equation}
e^{-\lambda(l)} \left(\frac{\kappa'(l)}{r} + \frac{1}{r^2} \right) - \frac{1}{r^2} = - F(r,l), 
\label{2ast}
\end{equation}
\begin{equation}
~e^{-\lambda(l)} \left(\frac{\lambda'(l)}{r} - \frac{1}{r^2} \right) + \frac{1}{r^2} = + F(r,l).
\label{3ast}
\end{equation}

To solve these equations, we recall that, far from the gravitating body, the gravitational field is weak and can be represented by its Newtonian limit. In the case of the de Sitter-invariant general relativity, the de Sitter-modified Newtonian potential is given by~\cite{DE}
\begin{equation}
\phi_{dS} = -\, GM \left( \frac{1}{r} + \frac{r}{2 l^2} \right). 
\label{SoludS2}
\end{equation}

Substituting into the Newtonian relation
\begin{equation}
g_{00} = 1 + 2 \, \frac{\phi_{dS}}{c^2},
\label{ge0012dS}
\end{equation}
we obtain an expression for the de Sitter-invariant Schwarzschild metric, 
\begin{equation}
g_{00} \equiv e^{-\lambda(l)} \equiv e^{\kappa(l)} = 1 + \beta \left(\frac{1}{r} +
\frac{r}{2 l^2} \right),
\label{g00dS}
\end{equation}
with $\beta$ an integration constant. Substituting this metric into Eq.~\eqref{3ast}, the function $F(r,l)$ is found to be
\begin{equation}
F(r,l) = \frac{\beta}{l^2 r}.
\label{4ast}
\end{equation}

The field equation~\eqref{3ast} with the right-hand side given by \eqref{4ast}, can then be easily integrated, yielding the de Sitter-invariant Schwarzschild metric~\eqref{g00dS}, with the integration constant given by:
\begin{equation}
\beta = - r_M \equiv - \frac{2 G M}{c^2}.
\end{equation} 

Using the above results, the de Sitter-invariant Schwarzschild solution can be written in the form:
\begin{equation}
ds^2 = f(r) \, c^2 \, dt^2 - 
\frac{dr^2}{f(r)} - r^2 \left(d\theta^2 + \sin^2\theta d\phi^2 \right)
\label{Schwar2dS}
\end{equation}
where\footnote{From now on, we use natural units, in which $\hbar = c = G = 1$.}
\begin{equation}
f(r) = 1 - \left(\frac{r_M}{r} + \frac{r}{r_\Lambda} \right),
\label{f(r)}
\end{equation}
with:
\begin{equation}
r_M = 2 M \quad \mbox{and} \qquad r_\Lambda = \frac{l^2}{M} \equiv \frac{2 l^2}{r_M}.
\label{horizons}
\end{equation}

These expressions show that the de Sitter-invariant black hole has two horizons: an event horizon with radius~$r_M$ and a cosmic horizon with radius~$r_\Lambda$. In this approach, the proper conformal current~$K^{\mu \nu}$ of ordinary matter is the source of the background de Sitter spacetime. Since $K^{\mu \nu}$ depends on the energy-momentum current $T^{\mu \nu}$, as can be seen from Eq.~\eqref{T&K}, the radius~$r_\Lambda$ of the cosmic horizon turns out to depend on the radius~$r_M$ of the event horizon.

\subsection{Asymptotic behavior}

The Newtonian limit of standard general relativity is obtained when the gravitational field is weak and the particle velocities are small. In the presence of a cosmological term $\Lambda$, some subtleties related to the process of group contraction show arise. For example, in the non-relativistic contraction limit $c \to \infty$, the Poincar\'e group reduces to the Galilei group. However, the Newton-Hooke group does not arise directly from the de Sitter group under the same contraction limit. The reason  is that, under such a limit, the boost transformations are lost. To obtain the physically relevant result, in which the group dimension is preserved, one has to simultaneously consider the limits $c \to \infty$ and $\Lambda \to 0$, but in such a way that
\begin{equation}
\lim c^2 \Lambda = \frac{1}{\tau^2} 
\end{equation}
with $\tau$ a finite time parameter. This means that, in the presence of $\Lambda$, the usual Newtonian limit must be supplemented with the small $\Lambda$ condition~\cite{gibbons}:
\begin{equation}
\Lambda r^2 \ll 1,
\end{equation}
which is equivalent to $r^2/l^2 \ll 1$. Consequently,
\begin{equation}
\lim_{r \to \infty} r^2/l^2 = 0 \qquad \mbox{and} \qquad \lim_{r \to \infty} r/l^2 = 0.
\label{limits0}
\end{equation}

As we have seen, the metric of the usual Schwarzschild solution reduces to the Minkowski metric at infinity. Using the limits~\eqref{limits0}, one sees that the de Sitter-invariant Schwarzschild metric also reduces to Minkowski at infinity. Such asymptotic behavior results from the cosmological term $\Lambda$ being non-constant in the de Sitter-invariant approach to gravitation. Recall that, in this approach, gravitation and $\Lambda$ are both sourced by ordinary matter. Since all source fields vanish at infinity, gravitation and $\Lambda$ will also vanish. This property is a crucial difference concerning standard general relativity, where the constancy of $\Lambda$ plagues the theory with divergences that obscure the physically relevant results~\cite{Susskind}.

\section{Black hole horizons}

The horizons $r_M$ and $r_\Lambda$, given by Eq.~\eqref{horizons}, are not manifest in the Sitter-invariant Schwarzschild metric~\eqref{Schwar2dS}. In fact, it is not singular for $r = r_M$ or $r = r_\Lambda$. Therefore, this metric is not appropriate for studying black hole thermodynamics.

To obtain a metric in which the horizons become manifest, one has to impose the condition $f(r) = 0$, which yields the polynomial equation
\begin{equation}
r^2 - r_\Lambda \, r + r_M \, r_\Lambda = 0.
\end{equation}

This polynomial equation has two real roots provided
\begin{equation}
r_\Lambda > 4 r_M.
\label{proviso}
\end{equation}

In this case, the two solutions, denoted $R_M$ and $R_\Lambda$, are found to be
\begin{equation}
R_M = \frac{r_\Lambda}{2} \left[1 - \Big(1 - \frac{4 \, r_M}{r_\Lambda} \Big)^{1/2} \right],
\label{1RM}
\end{equation}
and
\begin{equation}
R_\Lambda = \frac{r_\Lambda}{2} \left[1 + \Big(1 - \frac{4 \, r_M}{r_\Lambda} \Big)^{1/2} \right].
\label{1RL} 
\end{equation}

It follows from these expressions that\footnote{Note that condition~\eqref{NoNari} precludes the existence of a Nariai version of the de Sitter-invariant Schwarzschild solution.}
\begin{equation}
R_\Lambda > R_M.
\label{NoNari}
\end{equation}
In terms of $R_M$ and $R_\Lambda$, the original horizons are written as
\begin{equation}
r_M = \frac{R_\Lambda \, R_M}{R_\Lambda + R_M} \quad \mbox{and} \quad
r_\Lambda = R_\Lambda + R_M.
\label{naked horizons}
\end{equation}

Substituting into Eq.~\eqref{f(r)}, it takes the form
\begin{equation}
f(r) = \frac{(r - R_M) (R_\Lambda - r)}{(R_\Lambda + R_M) \, r},
\label{Phyf(r)}
\end{equation}
where the two horizons are now manifest. Furthermore, condition~\eqref{proviso} takes the form
\begin{equation}
(R_\Lambda + R_M)^2 > 4 R_M R_\Lambda.
\end{equation}

\subsection{Entropy}

The entropy associated with a given horizon is defined as $S_i = A_i/4$, with $A_i$ the area of the horizon. In the case of the event horizon, it is given by
\begin{equation}
S_M = \pi\, R_M^2.
\end{equation}

In the case of the cosmic horizon, it is given by
\begin{equation}
S_\Lambda = \pi\, R_\Lambda^2.
\end{equation}

Since $R_\Lambda > R_M$, then $S_\Lambda > S_M$. 

\subsection{Temperature}

The horizon temperature is defined as $T_i = {\kappa_i}/{2 \pi}$, where
\begin{equation}
\kappa_i = \frac{1}{2} \left|\frac{d f}{dr} \right|_{r=R_i}
\end{equation}
{is the surface gravity, with $f(r)$ given by Eq.~\eqref{Phyf(r)}. For the event horizon, the temperature~is}
\begin{equation}
T_M = \frac{1}{4 \pi} \left[\frac{R_\Lambda - R_M}{R_M (R_\Lambda + R_M)} \right].
\label{Shwarstempe}
\end{equation}

On the other hand, for the cosmic horizon, it has the form
\begin{equation}
T_\Lambda = \frac{1}{4 \pi} \left[\frac{R_\Lambda - R_M}{R_\Lambda (R_\Lambda + R_M)} \right].
\label{dStempe}
\end{equation}

Considering that $R_\Lambda > R_M$, it follows that $T_\Lambda < T_M$.

\section{First law of black hole thermodynamics}

In standard general relativity, the first law of black hole thermodynamics, which holds on the event horizon, is expressed as
\begin{equation}
d E_M = T_M\, dS_M.
\label{3Event}
\end{equation}
When there is a cosmological term~$\Lambda$ in the universe, like in the Schwarzschild-de Sitter  solution, an independent similar law of the form
\begin{equation}
d E_\Lambda = T_\Lambda\, dS_\Lambda,
\label{3dS}
\end{equation}
holds on the de Sitter cosmic horizon~\cite{GibHaw}.

In the case of de Sitter-invariant general relativity, the Schwarzschild solution has both an event and a cosmic horizon. The conserved energy density of the solution, given by the $(0 0)$ component of the Noether current~\eqref{Pi=T-K}, is
\begin{equation}
\varepsilon_{\Pi} = \varepsilon_M - \varepsilon_\Lambda,
\end{equation}
where $\varepsilon_M = T^{00}$ is the {\em translational} notion of energy density and $\varepsilon_\Lambda = K^{00}/4l^2$ is the {\em proper conformal} notion of energy density~\cite{dScosmo}. In terms of energy, it is written as
\begin{equation}
E_{\Pi} = E_M - E_\Lambda.
\label{Energy}
\end{equation}

Consequently, the corresponding first law of  black hole thermodynamics is written in the~form
\begin{equation}
d(E_M - E_\Lambda) = T_M dS_M - T_\Lambda dS_\Lambda.
\label{68}
\end{equation}

In this theory, therefore, the event and cosmic horizons are not independent but form a unique, entangled system. Since the cosmological term~$\Lambda$ is not constant, the dark energy $E_\Lambda$ is not constant either, and the thermodynamic variables can evolve along with cosmic time while preserving the thermodynamic equation~\eqref{68}. In the contraction limit $\Lambda \to 0$, corresponding to $l \to \infty$, it reduces to the thermodynamic equation~\eqref{3Event} of the Poincaré-invariant black hole.

\section{Black holes as reservoirs of dark energy}

In the de Sitter-invariant approach to gravitation, any source field gives rise to energy-momentum and proper conformal currents. Whereas the energy-momentum current appears as the source of the dynamic curvature of gravitation, the proper conformal current is the source of the kinematic curvature of the background de Sitter spacetime. Since the cosmological term~$\Lambda$ is an inherent part of the theory, all solutions to the de Sitter-invariant Einstein's equation~\eqref{NewModiEinstein} comprise a gravitational field and a local cosmological term~$\Lambda$. Consequently, the Sitter-invariant black hole carries matter ($E_M$) and dark ($E_\Lambda$) energies. 

However, although ubiquitous, the cosmological term will be relevant only for physical systems with high enough energy densities. This is the case of black holes, in which the dark energy associated with $\Lambda$ can represent a relevant part of the black hole energy. The amount of each kind of energy it carries is governed by the thermodynamic equation~\eqref{68} and can change over cosmic time.

\section{Cosmological coupling of black holes}

The de Sitter-invariant black hole features an event horizon and a cosmic horizon, forming a unique entangled system. These horizons show a fundamental difference: whereas the event horizon has a singularity at the center of the horizon, the cosmic horizon does not have~(see Eq.~\eqref{f(r)}. Consequently, whereas the event horizon is always tied to a definite position in space, the cosmic horizon is not. This position-indefiniteness allows us to assume that all black holes contribute to an effective cosmological term for the universe.

Thus, if the dark energy stored in all black holes can provide the necessary amount of energy to account for the dark energy we measure today, one can assume that the dark energy stored in black holes represents the universe's dark energy. Such an assumption establishes a link between the black hole's dynamics and the universe's evolution.

Of course, if a cosmological coupling of black holes exists, it should be at work at any cosmic time. In particular, it should hold in the primordial universe. In this case, to drive inflation and the subsequent universe expansion, it would be necessary to presuppose the existence of primordial black holes with $E_\Lambda \gg E_M$. This assumption represents a crucial step because, today, black holes are primarily associated with the gravitational collapse of old, giant stars.

\subsection{Observational evidences}

Recent cosmological observations~\cite{BHDE} found evidence of a {\em cosmological coupling of black holes}. To establish this coupling, one must assume that the gravitating mass of black holes increases with the universe's expansion, independent of accretions or mergers. Such a mass increase produces a concomitant increase in the dark energy carried by the black holes, thereby contributing to the dark energy density in Friedmann's equations. Furthermore, by considering the black hole production inferred from the history of cosmic star formation, it is concluded that black holes could account for the universe's total amount of dark energy, as measured today. As the remnants of stellar collapse, black holes are considered the origin of the universe's dark energy. Such a mechanism could explain, for example, the late-time acceleration of the universe's expansion.

Despite presenting some conceptual differences, the de Sitter-invariant approach to gravitation can be said to corroborate the experimental evidence of a cosmological coupling of black holes. There are, however, some crucial differences. For example, in the de Sitter-invariant approach to gravitation, the source of dark energy is not the black hole itself but the proper conformal current of ordinary matter. It is also unclear how the cosmological coupling of black holes would work in other cosmic periods, such as the primordial universe. Notwithstanding the differences, we believe this topic deserves further study, as establishing a cosmological coupling of black holes could answer crucial problems in modern cosmology.

\section{Discussion}

Unlike Poincaré-invariant special relativity, the de Sitter-invariant special relativity does include the proper conformal transformation in the spacetime kinematics.
The presence of these transformations has profound implications for de Sitter-invariant general relativity. For example, all solutions to the de Sitter-invariant Einstein's equation comprise a gravitational field, whose source is the energy-momentum current of ordinary matter, and a local cosmological term, sourced by the proper conformal current of the same ordinary matter~\cite{dScosmo}.

In particular, the de Sitter-invariant Schwarzschild solution represents a black hole with both an event and a de Sitter cosmic horizon. Consequently, besides lodging matter energy~$E_M$, black holes also lodge dark energy~$E_\Lambda$.
In contrast to the standard Schwarzschild black hole, which is specified by its mass~$M$ only, the de Sitter-invariant black hole is specified by its mass~$M$ and the local cosmological term~$\Lambda$, which can be interpreted as a hair of the black hole. Furthermore, since $\Lambda$ is encoded in the spacetime's local kinematics, which has a quantum nature by construction, it can be considered a quantum hair. Due to this additional structure, which originates from the proper conformal sector of the theory, the de Sitter-invariant approach yields an entirely new black hole solution. Possible implications of this solution for the physics of black holes, as well as for cosmology, will be discussed elsewhere.

%
%
%
%
\vspace{0.8cm} 
 
\noindent
{\bf Acknowledgments:} {The authors thank A. Coley and R. van den Hoogen for useful discussions. DFL thanks Dalhousie University, Canada, for a Ph.D. scholarship. SAN thanks Conselho Nacional de Desenvolvimento Cient\'{\i}fico e Tecnol\'ogico, Brazil, for a post-doctor fellowship (Contract No.~166099/2020-1). JGP thanks Conselho Nacional de Desenvolvimento Cient\'{\i}fico e Tecnol\'ogico, Brazil, for a research grant (Contract No.~312094/2021-3).}
\vskip6pt


\end{document}